\documentclass[aps,showpacs,superscriptadress,preprint]{revtex4}
\usepackage{bbm}
\usepackage{graphicx}
\usepackage{dcolumn}
\usepackage{bm}
\usepackage{here}
\usepackage{amsmath}
\begin{document}
\baselineskip=0.8 cm
\title{{\bf A preliminary analysis of the energy transfer between the dark
    sectors of the Universe}}

\author{Jia Zhou$^{1}$, Bin Wang$^{1}%
$\footnote{wangb@fudan.edu.cn }, Diego Pav\'{o}n
$^{2}$\footnote{diego.pavon@uab.es}, Elcio
Abdalla$^{3}$\footnote{eabdalla@fma.if.usp.br }}
\affiliation{$^{1}$ Department of Physics, Fudan University,
200433 Shanghai, China} \affiliation{$^{2}$ Department of Physics,
Autonomous University of Barcelona, 08193 Bellaterra, Barcelona,
Spain} \affiliation{$^{3}$ Instituto de F\'{\i}sica, Universidade
de S\~ao Paulo, CP 66318, 05315-970, S\~ao Paulo, Brazil.}

\vspace*{0.8cm}
\begin{abstract}
\baselineskip=0.6 cm We study the mutual interaction between the
dark sectors (dark matter and dark energy) of the Universe by
resorting to the extended thermodynamics of irreversible processes
and constrain the former with supernova type Ia data. As a
byproduct, the present dark matter temperature results in good
agreement with independent estimates of the temperature of the gas
of sterile neutrinos.

\end{abstract}
 \maketitle
\newpage

It is  widely held by now that the accelerated expansion of the
Universe can be traced to some invisible  agent (possibly the
quantum vacuum or some or other scalar field), dubbed dark energy
(DE), which takes slightly over 70\% of the energy budget and is
endowed with a high negative pressure -see \cite{reviews} for
reviews. The possibility that this agent interacts with the
remaining fields of the standard model is more natural and general
than otherwise whence it is receiving growing attention -see
\cite{korean-review} and references therein. However, local
gravity experiments set severe bounds on the interaction with
conventional matter (e.g., baryons) \cite{severe} but there is
nothing against a possible interaction with dark matter (DM).
Unless there exists an underlying symmetry that would set the
interaction (coupling) to zero (such a symmetry is still to be
discovered) there is no {\it a priori} reason to toss it away.
Moreover, the existence of this coupling is undeniable on physical
grounds: since DE energy gravitates it must be accreted by massive
compact objects like black holes and neutron stars. In a
cosmological context this energy transfer from DE to DM must be
small but non-vanishing.

The coupling between DE and dark matter (DM), first introduced to
lower down the huge value of the cosmological constant
\cite{wetterich}, can provide a mechanism to alleviate the
coincidence problem \cite{luca,winfried,german} or even solve it
\cite{solve}. Furthermore, it has been argued that an appropriate
interaction between DE and DM can influence the perturbation
dynamics and affect the lowest multipoles of the CMB spectrum
\cite{binw1,binw2}. Recently, it has been shown that such a
coupling could be inferred from the expansion history of the
Universe, as manifested in the supernova data together with CMB
and large-scale structure \cite{feng1,guo,bin1,bin2}. Signatures
of the interaction between DE and DM in the dynamics of galaxy
clusters has also been analyzed \cite{orfeu,elcio}. Further
discussions on the interaction between dark sectors can be found
in \cite{lin,tetradis}.

The aforesaid interaction has also been considered  from a
thermodynamical perspective -see e.g. \cite{bin3, bin4}. Assuming
that the DE can be treated as a fluid with a well defined
temperature, it has been argued that if at present there exists a
transfer of energy between DE and DM, it must be such that the
latter gains energy from the former and not the other way around
for the second law of thermodynamics \cite{callen} to be
fulfilled. The authors of Ref. \cite{bin4} considered a system
composed of two subsystems (DM and DE) at different temperatures.
In virtue of the extensive property, the entropy of the whole
system is the sum of the entropies of the individual subsystems
which (being equilibrium entropies) are just functions of the
energies of DE and DM even during the energy transfer process.

The aim of this Letter is to look more closely into the
aforementioned transfer between both subsystems by resorting the
thermodynamics of irreversible processes as formulated by Jou {\it
et al.}  \cite{david}. There, to cope with nonequilibrium (i.e.,
irreversible) situations,  the entropy is generalized by allowing
nonequilibrium quantities to enter it. Recall that, by definition,
nonequilibrium quantities vanish at equilibrium and, consequently,
this generalized entropy reduces to the conventional equilibrium
entropy in that limit.

If DE and DM conserve separately in an expanding
Friedmann-Lema\^{i}tre-Robertson-Walker universe we would have
\begin{eqnarray}\label{consv1a}
\dot{\rho}_m&+&3H(\rho_m+ p_{m})=0 \, , \\
\label{consv1b} \dot{\rho}_{x}&+&3H(1+w_{x})\rho_{x}=0\, ,
\end{eqnarray}
where the equation of state of DM can be approximately written in
parametric form as \cite{degroot}
\begin{equation}\label{matter-pressure}
\rho_{m} = n_{m} \, M + \textstyle{3\over{2}}\,  n_{m} \, T_{m}\,
,\quad \quad p_{m} = n_{m} \, T_{m} \qquad  (k_{B} = 1)
\end{equation}
provided that $T_{m} \ll M$. Therefore, $\rho_{m} \sim a^{-3}$ and
$\rho_{x} \propto \exp \int{-3(1+w_{x})\, da/a}$. Here, $w_{x} =
p_{x}/\rho_{x} < 0 $ is the equation of state parameter of dark
energy.

The dependence of both temperatures on the scale factor
\begin{equation}\label{tevol1}
T_m \propto a^{-2} \, , \qquad \qquad T_x \propto \exp \int{-3
w_{x}\, da/a}
\end{equation}
follows from integrating the evolution equation, $\dot{T}/T = -3 H
(\partial p/\partial \rho)_{n}$ for each subsystem. The latter is
a consequence of Gibbs' equation, $T dS = d(\rho/n) + p \, d(1/n)$
and the condition for $dS$ to be a differential expression. Since
$w_{x} <0$, equations (\ref{tevol1}) suggest that currently $T_{m}
\ll T_{x} $.

When the DE and DM components interact with each other, Eqs.
(\ref{consv1a}) and (\ref{consv1b}) generalize to
\begin{eqnarray}\label{consv2a}
\dot{\rho}_m&+&3H(\rho_m+ p_{m})= \Gamma \, , \\
{\rm and} \hfill &&\nonumber \\
\label{convs2b} \dot{\rho}_{x}&+&3H(1+w_{x})\rho_{x}= -\Gamma \, ,
\end{eqnarray}
respectively, where $\Gamma$ denotes the interaction term. For
$\Gamma
>0$ the energy proceeds from dark energy to dark matter.

For simplicity we assume the specific coupling
\begin{equation}\label{Q1}
 \Gamma = 3H \lambda \, \rho_{x} \, ,
\end{equation}
with $\lambda$ a small, dimensionless, positive quantity. This
kind of models has  been considered in the literature -see e.g.
\cite{bin2,jian-hua,wd}- and show compatibility with observation
\cite{ad}.

Upon neglecting the DM pressure, the ratio between the energy
densities, $r=\rho_{m}/\rho_{x}$, is seen to obey
\begin{equation}
\frac{dr}{dx}=3(\lambda+\lambda{r}+w_{x}r),
\end{equation}
where $x=\ln{a}$.  When $w_{x}$ and $\lambda$ are constants the
latter equation integrates to
\begin{equation}\label{r(x)}
r(x)=\frac{[r_{(eq)}(\lambda+w_x)+\lambda]\,
e^{3(\lambda+w_x)(x-x_{(eq)})}\, -\lambda}{\lambda+w_x}\, ,
\end{equation}
The subscript $(eq)$ indicates the value taken by the
corresponding quantity when DE and DM are in thermal equilibrium.

In the scaling regime, i.e., when the ratio $r$ stays constant,
the coincidence of the two dark components is understood
\cite{winfried,german}. Outside this  regime $r$ may be considered
piecewise constant, especially for $z \leq 20$. In particular,
$\lambda=-\frac{w_{x}r_{0}} {r_{0}+1}$ about the present time,
where $r_{0} \simeq 3/7$ is the current $r$ value.

\begin{figure}[tbp]
\includegraphics[width=12.0cm,height=9.0cm]{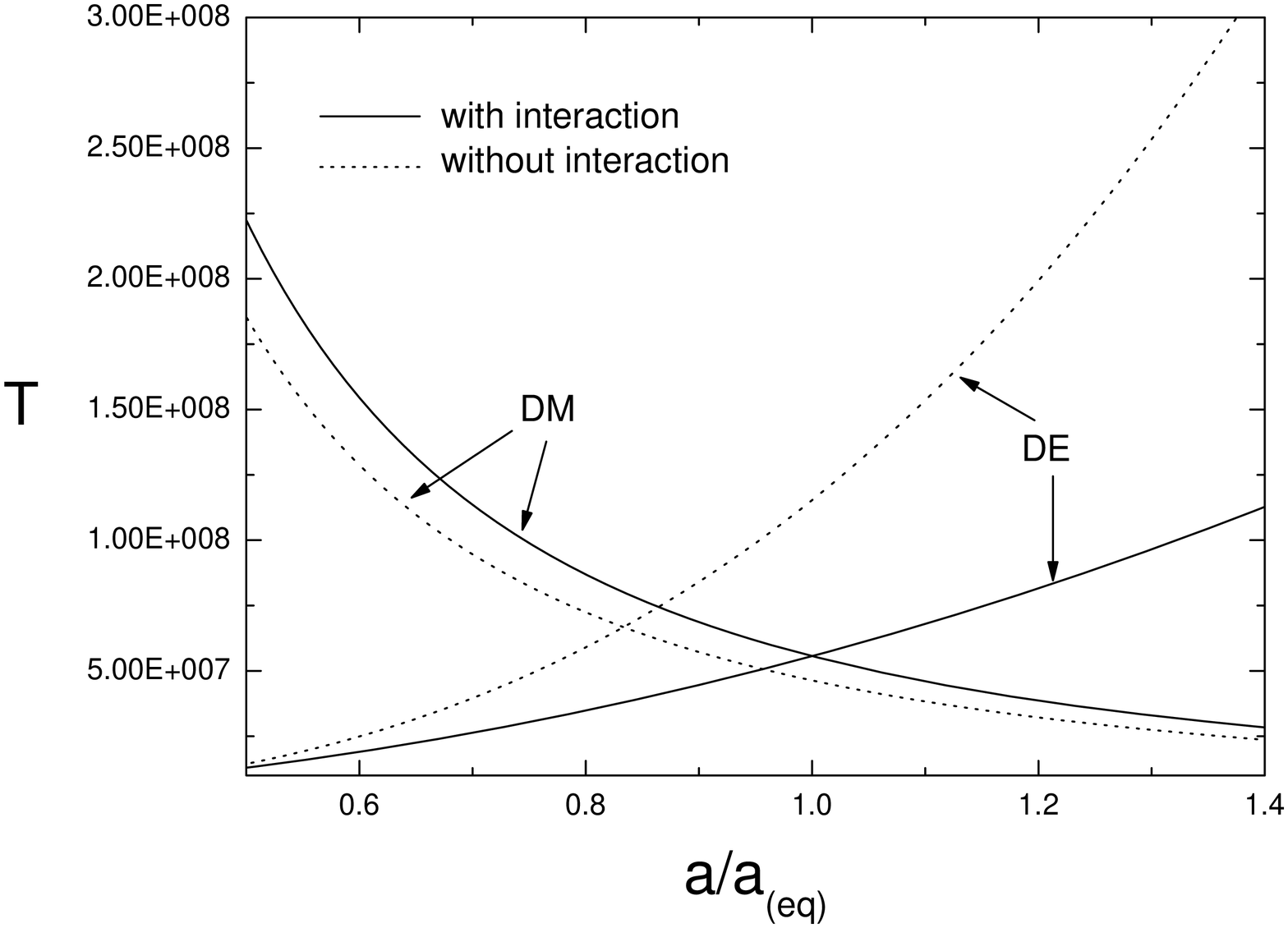}
\caption{Evolution of the temperatures (in Kelvin degrees) of DM
and DE in the non-interacting case (dotted lines), and in the
interacting case (solid lines). In drawing the figure we have
taken $w_{x} = -1$ and $\lambda = 0.3$.} \label{temp-evol}
\end{figure}

When the interaction is taken into account,  temperatures of DE
and DM evolve  as
\begin{equation}
T_{x}=T_{(eq)}\, e^{-3(w_{x}+\lambda)(x-x_{(eq)})}\, , \quad {\rm
and} \quad T_{m}= T_{(eq)} \, \frac{r}{r_{(eq)}} \,
e^{-[2+3(\lambda+w_{x})](x-x_{(eq)})} \, ,
\label{expressed}
\end{equation}
respectively. It is immediately seen that  both temperatures vary
more slowly than in the noninteracting case (recall that $0 <
\lambda < |w_{x}|$), see Fig.\ref{temp-evol}.

Consider an isolated system composed of two subsystems, one being
DE and the other DM,  satisfying
\begin{equation}
T_m\frac{dS_m}{dt}= \frac{dQ_m}{dt}=\frac{dE_m}{dt} \, , \qquad
{\rm and} \qquad
T_x\frac{dS_x}{dt}=\frac{dQ_{x}}{dt}=\frac{dE_{x}}{dt}+p_{x}\frac{dV}{dt}\,
, \label{satisfying}
\end{equation}
respectively,  with $E_m=\rho_{m} V$ , $E_{x}=\rho_{x} V$ and $V
= a^{3}$. Since the overall system is isolated, one has
\begin{equation}
\frac{dQ_m}{dt}=-\frac{dQ_x}{dt}=-\dot{Q} \, ,
\label{isolated}
\end{equation}
where $\dot{Q} = -a^{3}\, \Gamma$ is the  energy transfer rate. In
the comoving volume, this is nothing but the energy conservation
law:
$\dot{\rho}_{x}+\dot{\rho}_{m}+3H(\rho_{m}+\rho_{x}+p_{x})=0$.

In \cite{bin4}, the entropy of the whole system depends on the
energy densities and  volume only, and in virtue of the extensive
property, it is just the sum of the entropies, $S_{m}$ and
$S_{x}$, of the subsystems. In equilibrium thermodynamics
\cite{callen}, irreversible fluxes -such as energy transfers- play
no part and they do not enter the entropy function which is
defined for equilibrium states only. However, in nonequilibrium
extended thermodynamics such fluxes enter the entropy function,
$S^{*}$, which is more general than $S$ as it can be defined also
outside equilibrium \cite{david}. In this spirit, we postulate
that $S^{*} = S^{*}( \rho_{m}, \rho_{x}, V, \dot{Q})$.
Accordingly, its time variation is given by
\begin{equation}
\frac{dS^{*}}{dt}=T_{m}^{-1}\frac{dQ_{m}}{dt}+T_{x}^{-1}
\frac{dQ_{x}}{dt}-\tilde{\Gamma}(\dot{Q})\frac{d\dot{Q}}{dt} \, ,
\label{dSt}
\end{equation}
where $-\tilde{\Gamma}(\dot{Q})$ is defined as the derivative of
$S^*$ with respect to $\dot{Q}$ and we have used  Eqs.
(\ref{satisfying}). For the sake of simplicity, we suppose
$\tilde{\Gamma}(\dot{Q})=A \, \dot{Q}$, with $A$  a semi-positive
definite constant.

Assuming $r$ piecewise constant, the energy transfer rate at
present time can be roughly determined as
\begin{equation}
\left.\frac{d\dot{Q}}{dt}\right]_{0} \simeq -\frac{27 \, \lambda
\, a^{3}_{0} H^{3}_{0}}{8\pi (1+r_{0})}(\dot{H}+H^2)_{0} <0\, ,
\label{Qdot1}
\end{equation}
since the Universe entered the accelerated regime only recently.
Thus, in the current accelerating stage, the second law in
extended irreversible thermodynamics, $dS^*/dt\geq 0$, implies
\begin{equation}
A(d\dot{Q}/dt)\geq T_{x}^{-1}-T_{m}^{-1}\, .
\label{A1}
\end{equation}
In view of the evolution of DE and DM temperatures, this
inequality can be satisfied for any negative value of $A$.
However, for positive values, there must be an upper bound on $A$
to guarantee the second law.

In virtue of the expressions (\ref{Qdot1}) and (\ref{expressed}),
Eq. (\ref{A1}) can be recast around the present time as
\begin{equation}
-A'H^{3}\left( \frac{dH}{dx}+H \right)\geq e^{2.1x_{(eq)}}\,
e^{-5.1x}- \textstyle{3\over7}\, r_{(eq)}\, \left\{r_{(eq)}\,
e^{2x_{(eq)}}\, e^{x}\, +\,  e^{-0.1x_{(eq)}}\,
e^{3.1x}\right\}^{-1}, \label{Aprime}
\end{equation}
where $A'=567 \, T_{(eq)}\, A/(800\pi)$ and we have set $w_{x}$
and $r_{0}$ to $-1$ and $3/7$, respectively. Approximating the
Hubble function by $H \simeq H_{0}-H_{0} \, (1+q_{0})\, x$, with
$q_{0}$ the current value of the deceleration parameter,
integration  of the above inequality yields
\begin{equation}
H^{4}\geq H_{0}^{4}-\frac{4H_{0}^{4}}{5(1+q_{0})}+\frac{4 \,
 H_{0}^{4}}{5(1+q_{0})}[1-(1+q_{0})x]^{5}+
\frac{4e^{2.1x_{(eq)}}}{5.1\,
A'}(e^{-5.1x}-1)-\frac{4}{A'}\int_{x}^{0}\tilde{f}(x',r^{(eq)},x_{(eq)})dx'\,
, \label{H4}
\end{equation}
where
$\tilde{f}(x',r_{(eq)},x_{(eq)})=r_{(eq)}\{(r_{(eq)}-\textstyle{3\over
7})\, e^{2x_{(eq)}}e^{x}+  \textstyle{3\over 7}\,
e^{-0.1x_{(eq)}}e^{3.1x}\}^{-1}$. Using  $x=-\ln(1+z)$ we rewrite
the inequality as
\begin{eqnarray}
H^{4}(z) \geq
H_{0}^{4}&-&\frac{4H_{0}^{4}}{5(1+q_{0})}+
\frac{4H_{0}^{4}}{5(1+q_{0})}[1+(1+q_{0})\,
\ln(1+z)]^{5}\nonumber\\
&+&\frac{4(1+z_{(eq)})^{-2.1}}{5.1 \, A'}[(1+z)^{5.1}-1]-\frac{4}{A'}
\int_{0}^{z}\frac{\tilde{g}(z',z_{(eq)},r^{(eq)})}{1+z'}dz'
\, , \label{H4z}
\end{eqnarray}
where $\tilde{g}(z,z_{(eq)},r_{(eq)}) =
r_{(eq)}\{[r_{(eq)}-\textstyle{3\over 7}]\,
(1+z_{(eq)})^{-2}(1+z)^{-1}+ \textstyle{3\over 7}\,
(1+z_{(eq)})^{0.1}(1+z)^{-3.1}\}^{-1}$.

With the help of the expressions for the luminosity distance and
distance modulus,
\[
d_{L}= c \, (1+z)\int_{0}^{z}\frac{dz}{H(z)}\, , \quad {\rm and}
\quad \mu(z) = 5 \, \log \, (d_{L})+25 \, ,
\]
one obtains the following  inequality for the latter
\begin{equation}
\mu(z)\leq 5 \, \log \, \left[c\, (1+z) \,
\int_{0}^{z}h(z',A',z_{eq}, r_{(eq)})^{-\frac{1}{4}}dz'\right]+25
\, ,
\label{mu(z)}
\end{equation}
where $c$ is the speed of light and $h(z, A', z_{(eq)},r_{(eq)})$
stands for the right-hand-side of Eq. (\ref{H4z}).

The independent parameters $r_{(eq)}$, $z_{(eq)}$, and $A'$ can be
roughly appraised by resorting the second law and SN Ia data. Here
we shall use the  sample of the recent golden SNe Ia, compiled in
\cite{riess}, in a similar way as in \cite{zhou}. Since we have
employed a linear approximation in the above integration and
focused on the present accelerated era, we will use only low
redshift data, namely $z\leq 0.5$.

\begin{figure}[tbp]
\includegraphics[width=12.0cm,height=9.0cm]{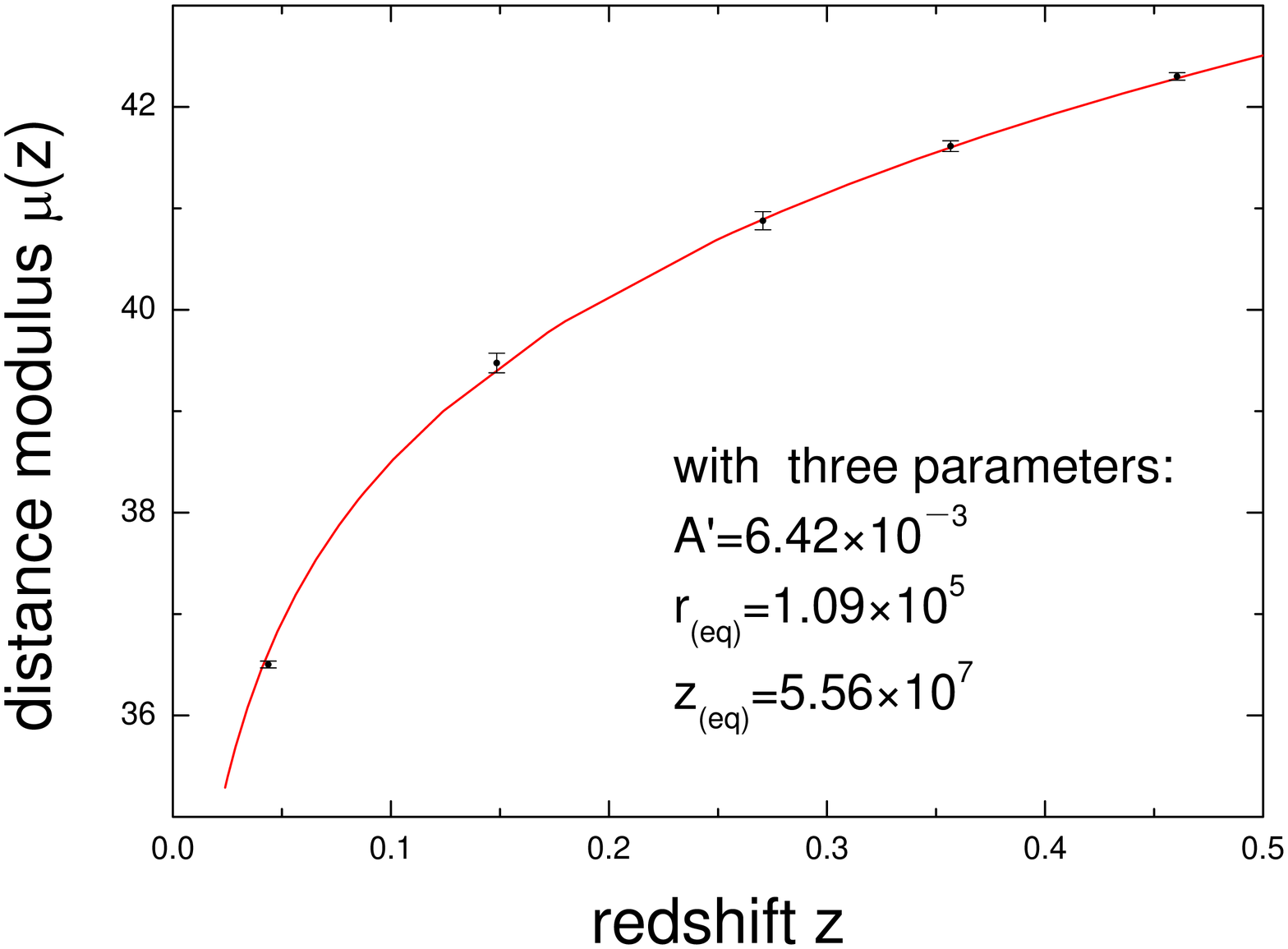}
\caption{The graph sets the border above which the second law of
thermodynamics would be violated.} \label{fig:mu(z)}
\end{figure}

Taking the equality sign in  (\ref{mu(z)}), setting $H_{0} = 72\, $
Km/s/Mpc, and  fitting the distance modulus to the supernovae data,
the best fit values of the parameters are found to be $A'=6.42
\times 10^{-3}$, $r_{(eq)}=1.09 \times 10^{5}$ and $z_{(eq)}=5.56
\times 10^{7}$, with $\chi^{2} = 73.6178102$. We remark that
$z_{(eq)}$ has only a small influence on the total $\chi^{2}$ value
when latter lies in the range $10^{3} \leq z_{(eq)} \leq 10^{8}$.
The value for $A'$ quoted above is the upper bound set by the second
law. For values exceeding that one the distance modulus curve
(depicted in Fig.\ref{fig:mu(z)}) would run below the SNe Ia data,
i.e., the second law, $dS^{*}/dt \geq 0$, would be violated in the
region above the curve.

As reasoned above (see also \cite{bin4}), after the thermal
equilibrium between DM and DE is lost (i.e., for $z < z_{(eq)}$),
a transfer of energy between these sectors will arise and $T_{x}$
will increase more slowly  than in the absence of interaction and,
correspondingly, $T_{m}$ will decrease more slowly -see Fig.
\ref{temp-evol}. This does not bring the systems to any new
equilibrium, but it certainly slows down the rate at which they
move away from mutual equilibrium. To assess quantitatively the
relevance of the interaction appearing in the expression for
$dS^{*}/dt$ we compute the current value of the ratio between the
third and first terms  in Eq. (\ref{dSt}), i.e.,
\begin{equation}
R_{m0}=\frac{|A\dot{Q}(d\dot{Q}/dt)|_0}{|T_{m}^{-1}(dQ_{m}/dt)|_0}
=\frac{11\, \alpha}{20\, r_{(eq)}}\left\{(r_{(eq)}-
\textstyle{3\over 7}\, e^{2x_{(eq)}}+ \textstyle{3\over 7}\,
e^{-0.1x_{(eq)}}\right\},
\label{Rm0}
\end{equation}
where $\alpha=A'\, H_{0}^{4}$. Inserting  the best fitting values
for $A'$, $r_{(eq)}$ and $z_{(eq)}$, we get $R_{m0} \simeq 1.4$.
Thus, the term coming from the interaction should not be neglected
in the expression for the entropy variation.

We next compare the present DM temperature with and without
coupling between the dark sectors. In the absence of interaction,
in virtue of Eq. (\ref{tevol1}), we would have
\begin{equation}
T_{m_0}=T_{(eq)}(1+z_{eq})^{-2} \simeq
\frac{T_{(eq)}}{z^2_{(eq)}}\, ,
\end{equation}
and the thermal equilibrium between DE and DM should have occurred
earlier (at larger $z_{(eq)}$) than in the interacting case.  If
we consider $T \sim z$ and use the best fitting value of
$z_{(eq)}$, we infer that $T_{m_0} \sim 10^{-7}$ Kelvin. In the
noninteracting case, the current DM temperature can be even
smaller. In the interacting case, the present temperature of DM
can be estimated as 
\begin{equation}
T_{m_0}=\frac{r_{0}}{r_{(eq)}}\, T_{eq}\, (1+z_{(eq)})^{0.1} \, .
\end{equation}
As noted above  $z_{(eq)}$ cannot be  strictly constrained since
$\chi^2$ shifts only by a factor of about $10^{-5}$ when
$z_{(eq)}$ varies from $10^3$ to $10^8$. For $(\alpha, r_{(eq)},
z_{eq})=(1.64\times 10^5, 0.8\times 10^5,
\newline 3.87\times 10^4)$ --- what implies $\chi^{2} = 73.61782324$,
and we obtain $T_{m_0} \simeq 0.596$ Kelvin. This value is
consistent with the estimations of Hansen {\em et al.}
\cite{hansen} based on simulations of the number of satellite
galaxies and astronomical upper limit of the sterile neutrinos
-one of the best promising DM candidates- with temperature
\begin{equation}
T_{w_0}=T_{\nu_0}\left(\Omega_w h^2\frac{\, 94\, {\rm
eV}}{m_w}\right)^{1/3}\simeq 0.581 \, {\rm Kelvin}
\end{equation}
(see Eq. (1) in Ref. \cite{hansen}), where $T_{\nu_0}= 1.946$
Kelvin, $\Omega_{w} = 0.3$, $h=0.72$ and the lower bound
$m_w=550\,$ eV \cite{viel}.

In summary, we have further employed the second law of
thermodynamics to study the coupling between the dark sectors of
the Universe. More specifically, by using the nonequilibrium
entropy, $S^{*}$, of extended irreversible thermodynamics
alongside  the gold SNe Ia data of Riess {\it et al.} \cite{riess}
we roughly determined the redshift, $z_{(eq)}$, at which DE and DM
were momentarily in thermal equilibrium as well as the ratio
between their energy densities at that moment. In addition, we
estimated an upper bound on the phenomenological parameter $A'$.
We would like to emphasize that in the absence of any mutual
interaction between dark sectors the DM temperature results
extremely low. However, due to the interaction it can meet the
independent astronomical upper bound on warm DM particles
\cite{hansen}.

Admittedly, our analysis is just preliminary and presents some
limitations. In the first place, we did not specify
microscopically the process by which the two dark sectors couple
with one another -this lies far beyond the scope of this work. We
only proposed a phenomenological expression for the interaction
(cf. Eq. (\ref{Q1})). Secondly, we expanded
$\tilde{\Gamma}(\dot{Q})$ just to first order in $\dot{Q}$. A more
ambitious treatment should go to higher orders. However, this
would introduce  additional unknown parameters and complicate the
numerical analysis, something to be concerned with in view of the
scarcity and limitations of current observational data.

This said, we believe our study may well open the way to fuller
investigations of the DE/DM interaction in the light of
nonequilibrium thermodynamics.

\acknowledgments{This work was partially supported by NNSF of
China, Shanghai Education Commission and Shanghai Science and
Technology Commission. D.P. is indebted to the Physics Department
of Fudan University for warm hospitality and financial support.
This work was partially supported by the ``Ministerio Espa\~{n}ol
de Educaci\'{o}n y Ciencia" under Grant No. FIS2006-12296-C02-01.
E.A.'s work was supported by FAPESP and CNPQ of Brazil.}

\end{document}